# $^{57}$Fe enrichment in mice for β-thalassaemia studies via Mössbauer spectroscopy of blood samples

George Charitou[*,1], Charalambos Tsertos[1], Yannis Parpottas[2], Marina Kleanthous[3], Carsten W Lederer[3], Marios Phylactides[3]

[1] *Department of Physics, University of Cyprus, 1678 Nicosia, Cyprus*

[2] *School of Engineering and Applied Sciences, Frederick University, 1036 Nicosia, Cyprus*

[3] *Molecular Genetics Thalassaemia Department, The Cyprus Institute of Neurology and Genetics, 1683 Nicosia, Cyprus*

**(Revised version: 03/04/2019)**

**Abstract:** In this work, wild-type and heterozygous β-thalassaemic mice were enriched with $^{57}$Fe through gastrointestinal absorption to characterize in more details the iron complexes appeared in the measured Mössbauer spectra. The $^{57}$Fe enrichment method was validated and Mössbauer spectra were obtained at 80K from blood samples from wild-type and β-thalassaemic mice at 1, 3, 6, and 9 months of age. As expected, the haemoglobin levels of the thalassaemic mice were lower than from normal mice indicating anaemia. Furthermore, significant amounts of ferritin-like iron were observed in the thalassaemic mice samples, which decreased with mouse age, reflecting the pattern of reticulocyte count reduction reported in the literature.

## Introduction

Beta-thalassaemia is a hereditary single-gene disorder characterized by decreased synthesis or absence of β-globin chains which prevents the production of normal levels of adult haemoglobin (HbA) and consequently the effective production of red blood cells (RBCs). It is a disease associated with considerable morbidity and mortality. Thalassaemia intermedia patients suffer from mild to moderate anaemia, prominent splenomegaly and bone deformities. Thalassaemia major patients, on the other hand, exhibit more severe symptoms including severe anaemia thus requiring regular blood transfusions from infancy. Moreover, iron overload occurs in thalassaemic patients due to increased gastrointestinal absorption and due to the required frequent blood transfusions (Eleftheriou 2003; Galanello and Origa 2010; Cappellini et al. 2014).

Haemoglobin, found inside the RBCs, is an essential globular metalloprotein which mainly carries oxygen from the lungs to the cells. Haemoglobin is comprised of two pairs of identical α and β globin chains (subunits). Each subunit consists of a protein chain bound to a haem moiety. Haem contains an iron ion in its centre, which can change between the ferrous (Fe$^{II}$) and ferric (Fe$^{III}$) state, depending upon the ligand it is attached to (Maeda 1979).

---

[*] Part of his PhD thesis

RBCs are produced through the process of erythropoiesis in the bone marrow, and in cases of extreme erythropoietic stress in the liver and spleen. During this multistep process, haematopoietic stem cells proliferate and mature into reticulocytes, which are released into the blood stream. After 1-2 days in the circulation, the reticulocytes differentiate further into mature RBCs. However, under some circumstances, such as sideroblastic anaemia, the developing erythroid cells can continue to take up iron beyond the cells physiological requirements (Ponka et al. 2013).

Despite its important role within the body, iron has the potential to damage tissues by catalyzing the conversion of hydrogen peroxide to free-radical ions. Therefore, iron homeostasis is tightly regulated, and mechanisms that control iron recycling, transport and storage exist. Iron is transported through the plasma bound to transferrin, a glycoprotein that can tightly bind two atoms of $Fe^{III}$ (Gkouvatsos et al. 2012), and it is primarily stored by ferritin in the liver, spleen and bone marrow. Ferritin is an iron binding protein with a spherical shell composed of 24 subunits of two different types and an 8 nm central cavity that can hold up to 4500 $Fe^{III}$ oxy-hydroxide atoms (Harrison and Arosio 1996; Knovich et al. 2009). Ferritin is also found extracellularly in the serum, which is used as a clinical marker of iron status (Knovich et al. 2009).

In various pathological conditions, such as haemochromatosis and thalassaemia, when transferrin saturation is elevated, iron can exist in the serum not bound to transferrin or any other traditional iron binding proteins (Brissot et al. 2012; Patel and Ramavataram 2012). This free iron is called non-transferrin bound iron (NTBI) and can catalyze the Fenton and Haber-Weiss reaction, producing free reactive radicals (Patel and Ramavataram 2012). NTBI represents a potential toxic iron form (Brissot et al. 2012).

$^{57}$Fe Mössbauer spectroscopy (MS) (Gütlich et al. 2011) is capable of characterizing iron complexes and in particular of providing information such as the iron electronic structure, magnetic structure, hyperfine interactions, valence/spin state, local microenvironment/iron stereochemistry and symmetry of environment, iron bonding, number of resonant nuclei, dynamics. This is in contrast to chemical or clinical markers which are able to detect only specific iron complexes.

Mössbauer spectroscopy was previously utilized to study iron complexes in blood and organs of humans and some animals. Two doublets were used to describe the oxyhaemoglobin in MS spectra of blood samples from human and animals (Hoy et al. 1986; Oshtrakh and Semionkin 1986; Oshtrakh et al. 2010). When spectra from RCBs of healthy people and patients with thalassaemia were compared, a new component was identified in the patient spectra which was considered to be ferritin-like iron (Xuanhui et al. 1988; Abreu et al. 1989; Jiang et al. 1994). Furthermore, in spectra of RCBs from patients, who underwent long-term therapy by regular blood transfusion and deferoxamine (an iron chelator), no ferritin was found in their RBCs, whereas ferritin was still present in the serum (Jiang et al. 1994).

In our previous work (Charitou et al. 2018), we obtained MS spectra from blood, liver and spleen samples from a normal mouse (wild-type C57BL/6) and a β-thalassaemic (th3/+) mouse. Thalassaemic blood samples showed evidence of anaemia while the thalassaemic liver and spleen samples revealed an increased deposition in ferritin-iron. The results demonstrated that the thalassaemic mouse model is a good candidate to study thalassaemia in more details by means of Mössbauer spectroscopy. In contrast to human patients, the thalassaemic mice did not receive any medical treatment, thus the iron components present in the samples were not altered.

In order to overcome the experimental difficulty of the low iron absorption in the MS spectra what was encountered by Charitou et al. (2018), mice were enriched with $^{57}$Fe through the gastrointestinal absorption. This should enable us to study in more detail the iron complexes present in these spectra. In this paper, we present the methodology of the enrichment procedure applied as well as a comparison between the spectra of blood samples from wild-type and thalassaemic mice at 1, 3, 6 and 9 months of age.

## Materials and methods

### Experimental Setup

The experimental setup, calibration procedure and production of the sample holders were previously described in detail by Charitou et al.(2018). The spectra were acquired at a temperature of 80±0.5K. The maximum velocity was set at ~4.10mm/s. All MS parameters extracted are given relative to metallic iron.

### Mouse Model

As a control group, wild-type C57BL/6 mice were used. As a β-thalassaemia model, heterozygous mice with deleted the b1 and b2 adult mouse globins (th3/+) were used, since homozygous (th3/th3) mice die perinatally (Yang et al. 1995). The th3/+ mice are bred on a C57BL/6 background. The th3/+ mice appear normal but show haematologic indices characteristics of severe thalassaemia, exhibit tissue and organ damage, which is typical of the disease, and also show spontaneous iron overload in the spleen, liver and kidneys.

### $^{57}$Fe enriched feed preparation

The increased $^{57}$Fe/Fe ratio in the mice was achieved through the continuous gastrointestinal absorption of a $^{57}$Fe-enriched diet. This absorption procedure is similar to the way that non-transfused thalassaemia intermedia patients develop iron overload over time. The $^{57}$Fe enriched feed was prepared by supplementing iron-deficient feed (Ssniff E15510 <10mg Fe/kg) with $^{57}$FeSO$_4$·7H$_2$O, which is a widely used iron salt for anaemia treatment.

A solution of the iron salt was prepared by dissolving 200 mg of $^{57}$Fe metallic iron (>95%, IsoFLEX USA), into 6 ml of 1M sulphuric acid (H$_2$SO$_4$) at ~ 40$^o$C. This solution had a characteristic bluish-green colour, thus indicating the existence of FeSO$_4$·7H$_2$O. To achieve an iron concentration of 20 mg/ml, 4 ml of distilled water were further added to the solution, and then stored at 4$^o$C.

According to the literature, the minimum requirements for iron in the mouse diet is 35 mg Fe/kg of feed while the optimum quantities are about 120 mg Fe/kg (National Research Council 1995). Due to the high price of the $^{57}$Fe isotope, we prepared mouse feed at 40-50 mg Fe/kg. For this purpose, 2 ml of the stored solution, which contains 40 mg $^{57}$Fe, were further diluted into 650 ml of sterilized water to ensure uniform distribution before thoroughly mixing with 1 kg of the powdered iron-deficient feed (Ssniff E15510). The mixture was then cut into small portions, left to dry and was eventually frozen at -18$^o$ C, until needed. This mouse feed corresponds to an $^{57}$Fe/Fe ratio of over 80%, which is much higher than in naturally occurring iron (2.12%).

### Sample population - Mouse breeding

Wild-type and th3/+ mice were raised and utilized for this study at 1, 3, 6 and 9 months of age. One mouse was raised in standard conditions with a typical diet, and sacrificed at 1 month of age for comparative purposes, to evaluate the enrichment method.

For the enrichment, six wild-type female mice were fed for a month with the Ssniff E15510 iron deficient diet to reduce the amount of natural iron in their bodies. Next, they were fed with the prepared $^{57}$Fe enriched diet for a week to increase their iron levels back to normal in order to be ready for a healthy pregnancy. For mating, these females were separated into pairs and a th3/+ male mouse was made available for each female pair. The females continued receiving the $^{57}$Fe enriched diet for the whole duration of mating, pregnancy and nurturing of the pups. After weaning, the pups were identified as either wild-type or th3/+ (group), based on their reticulocyte count levels using a haematology analyser (Sysmex XT2000 iVet, Sysmex Europe GmbH). Mice with >25% reticulocytes were considered thalassaemic while mice with <15% reticulocytes were considered normal. The pups continued receiving the $^{57}$Fe enriched feed throughout their lives while they had an unlimited access to fresh food and water.

**Sample Collection**

The blood from the mice was collected into sodium-heparin tubes to inhibit clotting. The blood volume collected from each mouse was 0.5-1 ml, therefore, the RBCs were not separated from plasma since the separation method (multiple centrifugations and washes) would result in the loss of sample volume and consequently, would result in less absorption. The samples where then placed into a MS holder of 10-mm in diameter.

The organs (liver, spleen) needed for the enrichment evaluation, were isolated after sacrificing the animals, and washed in 1x phosphate buffered saline (PBS) several times to remove as much excess blood as possible. The organs were then placed in MS holders.

The blood and organ samples in the holders were frozen in liquid nitrogen and stored until analysed by Mössbauer spectroscopy.

**Results and Discussion**

**Validation of $^{57}$Fe-enrichment method**

To distinguish the components of the $^{57}$Fe stored solution described previously and confirm the existence of FeSO$_4$·7H$_2$O, (a) a sample of powdered FeSO$_4$ (Sigma-Aldrich), dissolved in water, and (b) a sample from the $^{57}$Fe stored solution, which was further diluted with distilled water, were measured at 80K. The MS spectrum of the latter sample is shown in figure 1. Both spectra were fitted with a doublet and they exhibited similar MS parameters, as shown in Table 1.

**Table 1** The Mössbauer fitting parameters (Γ: FWHM, δ: isomer shift, ΔEq: quadrupole splitting) of the measured spectra at 80K of the $^{57}$Fe stored solution and the commercially obtained FeSO$_4$, prepared as described in the text. The overall instrumental error is < 0.03 mm/s. The errors shown are of statistical origin

|  | δ (mm/s) | ΔEq (mm/s) | Γ (mm/s) |
|---|---|---|---|
| FeSO$_4$ | 1.38±0.01 | 3.36±0.01 | 0.40±0.01 |
| $^{57}$Fe stored solution | 1.38±0.01 | 3.38±0.01 | 0.42±0.01 |

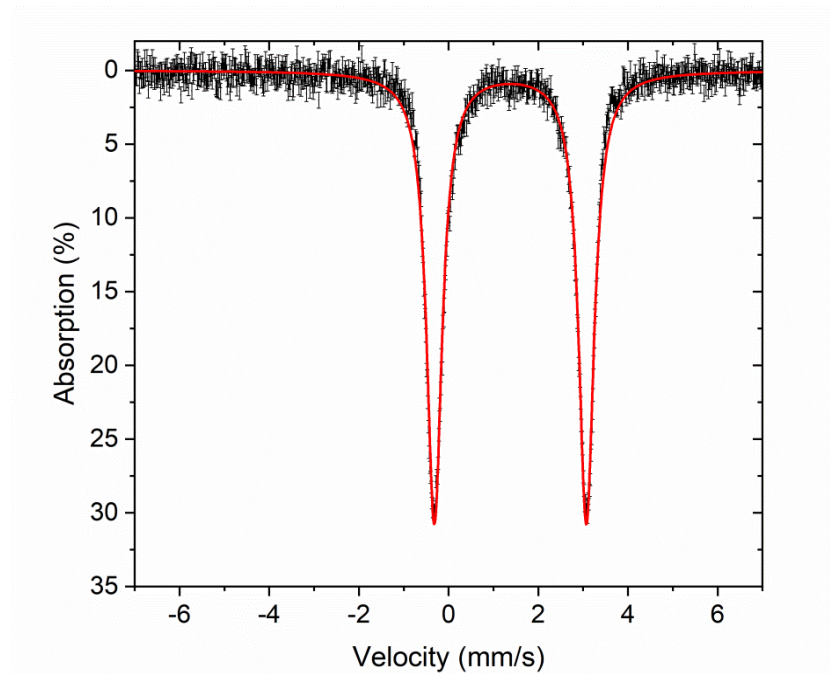

**Fig. 1** Mössbauer spectrum at 80K of the $^{57}$Fe stored solution, prepared as described in the text

Blood, liver and spleen samples from two normal mice at 1 month of age, with and without $^{57}$Fe enrichment, were analysed using MS to evaluate the effectiveness of the $^{57}$Fe enrichment method. Note that the pups had been weaned around 21-26 days after birth, therefore, most of the $^{57}$Fe identified in their tissues originated from maternal milk in their diet. Figure 2 shows the corresponding MS spectra. It can be clearly seen that the iron absorption is significantly higher in the samples derived from animals on the $^{57}$Fe enriched diet, resulting in spectra with a distinctly improved signal-to-noise (S/N) ratio that is adequate for a detailed quantitative analysis.

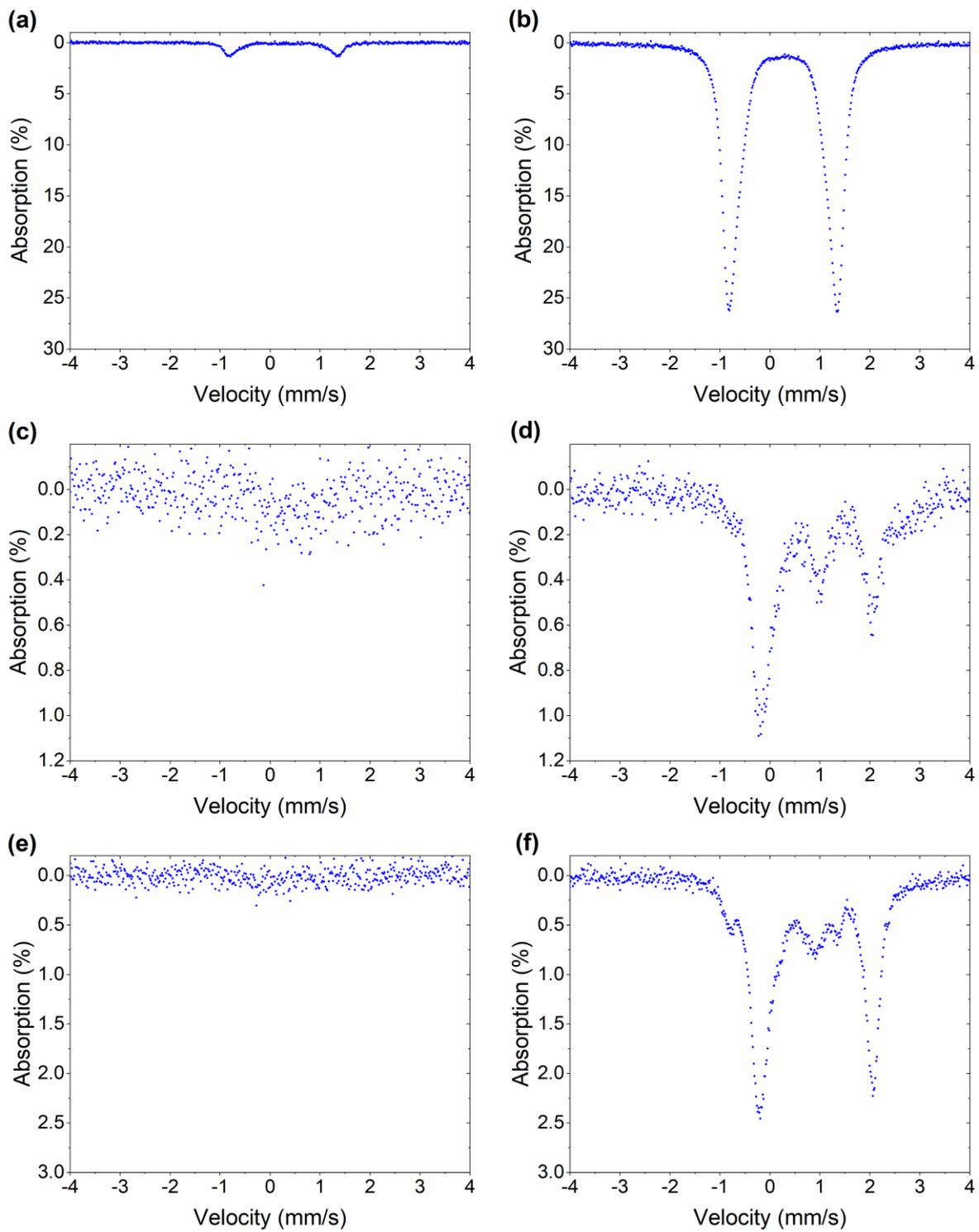

**Fig. 2** Mössbauer spectra from blood (a,b), liver (c,d) and spleen (e,f) of wild-type mice at 1 month of age, without (a, c, e) and with (b, d, f) $^{57}$Fe enrichment respectively. The scale is the same for each pair of spectra to show more clearly the increase of absorption due to the $^{57}$Fe enrichment

**Blood samples**

Blood samples from wild-type and th3/+ $^{57}$Fe enriched mice at 1, 3, 6 and 9 months of age (one mouse per age and group) were studied by means of MS. Figure 3 shows the measured and fitted MS spectra from the blood samples of wild-type and th3/+ $^{57}$Fe enriched mice at 1 and 9 months of age. In all spectra, a major doublet that exhibits the parameters of oxyhaemoglobin (oxy-Hb) is observed. This doublet was fitted with two different sub-doublets indicating the α- and β-globins of haemoglobin's tetramer (denoted as Hb-α and Hb-β, respectively). Note that, in other studies, two doublets were also used to describe the oxy-haemoglobin in spectra from blood samples of humans and animals (Hoy et al. 1986; Oshtrakh and Semionkin 1986; Oshtrakh et al. 2010).

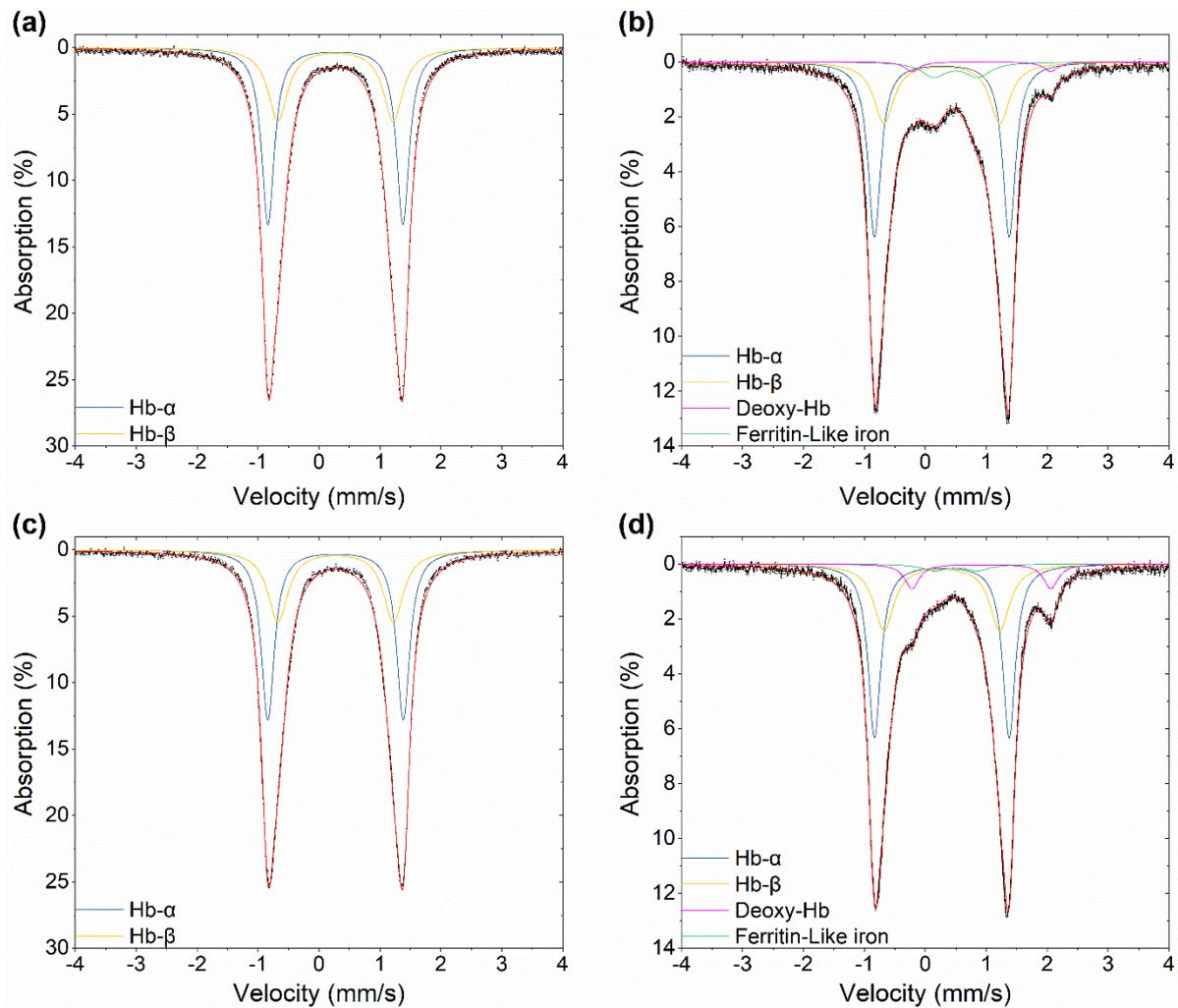

**Fig. 3** Mössbauer spectra obtained at 80K from the blood samples of mice at 1 (a, b) and 9 (c, d) months of age. Spectra (a) and (c) correspond to wild-type mice and spectra (b) and (d) to th3/+mice. All mice were $^{57}$Fe enriched

**Table 2** The Mössbauer fitting parameters (sub-doublet, relative percentage of the sub-doublet in the spectrum, Γ: FWHM, δ: isomer shift, ΔEq: quadrupole splitting, and the area of the sub-doublet) from the blood sample spectra of wild-type and thalassaemic mice at 1, 3, 6 and 9 months of age. The instrumental error is < 0.03 mm/s. The statistical errors are also shown. Note that no errors are reported for the parameters fixed in the fitting procedure

| Sample | Age (Months) | Sub-Doublet | Relative Percentage | Γ (mm/s) | δ (mm/s) | ΔEq (mm/s) | Area (x10⁻³) | Hb-α / Hb-β |
|---|---|---|---|---|---|---|---|---|
| **Wild-type** | 1 | Hb-α | 50 | 0.25±0.01 | 0.27±0.01 | 2.22±0.01 | 14.97±0.02 | 1 |
| | | Hb-β | 50 | 0.39±0.01 | 0.27±0.01 | 1.89±0.01 | 14.97±0.02 | |
| | 3 | Hb-α | 50 | 0.25±0.01 | 0.27±0.01 | 2.21±0.01 | 17.06±0.02 | 1 |
| | | Hb-β | 50 | 0.40±0.01 | 0.27±0.01 | 1.89±0.01 | 17.06±0.02 | |
| | 6 | Hb-α | 50 | 0.25±0.01 | 0.27±0.01 | 2.21±0.01 | 10.97±0.02 | 1 |
| | | Hb-β | 50 | 0.39±0.01 | 0.27±0.01 | 1.90±0.01 | 10.97±0.02 | |
| | 9 | Hb-α | 50 | 0.26±0.01 | 0.27±0.01 | 2.22±0.01 | 14.70±0.02 | 1 |
| | | Hb-β | 50 | 0.40±0.01 | 0.27±0.01 | 1.90±0.01 | 14.70±0.02 | |
| | Average | Hb-α | | 0.25 | 0.27 | 2.21 | | |
| | | Hb-β | | 0.40 | 0.27 | 1.90 | | |
| **Th3/+** | 1 | Hb-α | 43.8 | 0.25 | 0.27 | 2.21 | 7.13±0.02 | 1.12±0.01 |
| | | Hb-β | 39.1 | 0.40 | 0.27 | 1.90 | 6.36±0.04 | |
| | | Deoxy-Hb | 3.2 | 0.29 | 0.92 | 2.28 | 2.27±0.02 | |
| | | Ferritin-like iron | 13.9 | 0.49 | 0.50 | 0.72 | 0.52±0.02 | |
| | 3 | Hb-α | 45.0 | 0.25 | 0.27 | 2.21 | 9.39±0.04 | 1.01±0.01 |
| | | Hb-β | 44.5 | 0.40 | 0.27 | 1.90 | 9.30±0.05 | |
| | | Deoxy-Hb | 1.1 | 0.29 | 0.92 | 2.28 | 0.23±0.03 | |
| | | Ferritin-like iron | 9.4 | 0.49 | 0.50 | 0.72 | 1.97±0.04 | |
| | 6 | Hb-α | 49.6 | 0.25 | 0.27 | 2.21 | 7.50±0.03 | 1.16±0.01 |
| | | Hb-β | 43.0 | 0.40 | 0.27 | 1.90 | 6.49±0.04 | |
| | | Deoxy-Hb | 0 | 0.29 | 0.92 | 2.28 | 0 | |
| | | Ferritin-like iron | 7.4 | 0.49 | 0.50 | 0.72 | 1.12±0.04 | |
| | 9 | Hb-α | 43.5 | 0.25 | 0.27 | 2.21 | 6.98±0.02 | 1.03±0.01 |
| | | Hb-β | 42.2 | 0.40 | 0.27 | 1.90 | 6.78±0.03 | |
| | | Deoxy-Hb | 8.3 | 0.29 | 0.92 | 2.28 | 1.33±0.03 | |
| | | Ferritin-like iron | 5.9 | 0.49 | 0.50 | 0.72 | 0.96±0.02 | |

The fitting and analysis procedure of the measured MS spectra is rather complex and requires more iteration steps and different parameter fitting strategies to achieve final results consistent for all samples studied. To be more precise, at an initial attempt to fit the spectra of the wild-type mice by two Lorentzian sub-doublets, letting all other MS parameters free to vary, resulted in an Hb-α /Hb-β ratio distinctly higher than that of 51/49 found in our previous work (Charitou et al. 2018). Note that a Hb-α/Hb-β ratio of 1 is expected from biochemical studies (Yang et al. 1995; Schechter 2008). Deviations from the expected 1:1 ratio have been also discussed in the literature (Oshtrakh et al. 2010), where this ratio was finally fixed to 1. Thus, to keep our results consistent with the Hb-α/Hb-β ratio of 1, the following three steps were used in the fitting procedure.

I. At first, the spectra from the wild-type mice were fitted using two Lorentzian sub-doublets with a fixed 1:1 ratio, from which the average values for the corresponding Γ, δ and ΔEq MS parameters were extracted and presented in Table 2. These parameters are in agreement with our previous measurements (Charitou et al. 2018).
II. As can be seen in figures 3(b, d), the spectra from the thalassaemic mice are more complex, consisting of up to four overlapping sub-doublets. Therefore, the spectra from the th3/+ mice were simultaneously fitted by (i) the Hb-α and Hb-β sub-doublets fixed to the average values

- of step 1 (presented in Table 2) and (ii) by two additional sub-doublets with free varying parameters. From the results of this fit, the two additional sub-doublets were attributed to deoxy-haemoglobin and ferritin-like iron.
III. To eliminate any variations in the fit due to the strong overlapping, and to achieve maximum consistency between the spectra analysis, the thalassaemic spectra were refitted having all doublets fixed with only their area free to vary. The two oxy-haemoglobin's doublets were fixed to the fitting values extracted from the spectra of the wild-type mice in step I. The deoxy-haemoglobin's doublet was fixed to the values $\Gamma=0.29$ mm/s, $\delta=0.92$ mm/s and $\Delta E_q=2.28$ mm/s, as extracted from the corresponding fit of the spectra of the th3/+ mice at 9 month of age only, since the contribution of the corresponding doublets in the spectra of the other mice ages were too small to obtain reliable values. The ferritin-like iron's doublet was fixed to the values $\Gamma=0.49$ mm/s, $\delta=0.50$ mm/s and $\Delta E_q=0.72$ mm/s, which resulted from the average fitting values extracted in step II. It should be noted here that the values used for deoxy-haemoglobin and ferritin-like iron are in line with the results available in current literature (Kaufman et al. 1980; Hoy et al. 1986; Jiang et al. 1994; Chua-anusorn et al. 1999; Oshtrakh et al. 2011; Charitou et al. 2018).

The final results of the extracted MS parameters are summarised in Table 2, for the two mice species and the four mice ages studied. The Hb-α/Hb-β ratio of the th3/+ mice was found slightly elevated, with values between ~1 and 1.16. The latter increased ratio in the spectra of the thalassaemic mice might be due to an excess of α-globin chains associated with β-thalassaemia. The deoxyhaemoglobin, which is haemoglobin without any attached ligand normally present in the venous blood, found in three thalassaemic samples, (th3/+ mice at 1, 3 and 9 months of age) was probably due to some amount of deoxyhaemoglobin that did not get oxygenated during the sample collection and preparation process. On an average, for all ages of the mouse sample examined, the th3/+ mice had about 45% less total haemoglobin (fitting area of $Hb_{total}$ = Hb-α + Hb-β + deoxy-Hb = 0.16) relative to the wild-type ones (fitting area of $Hb_{total}$ = 0.29). This is in agreement with the values reported in the literature, as wild-type C57BL/6 mice have Hb levels of ~16g/dl (Physiological Data Summary – C57BL/6J 000664) and th3/+ mice 8-9g/dl (Gardenghi et al. 2007). The reduced haemoglobin levels are an indication of severe anaemia.

Figure 4 shows the fitted area of Hb in the spectra from the wild-type and th3/+ mouse samples, for all ages investigated. Hb was found to be unaffected by mouse age. This was expected for the samples from the wild-type mice, but not for the thalassaemic ones, since in most human thalassaemia intermedia patients the anaemia worsens, later on, in life.

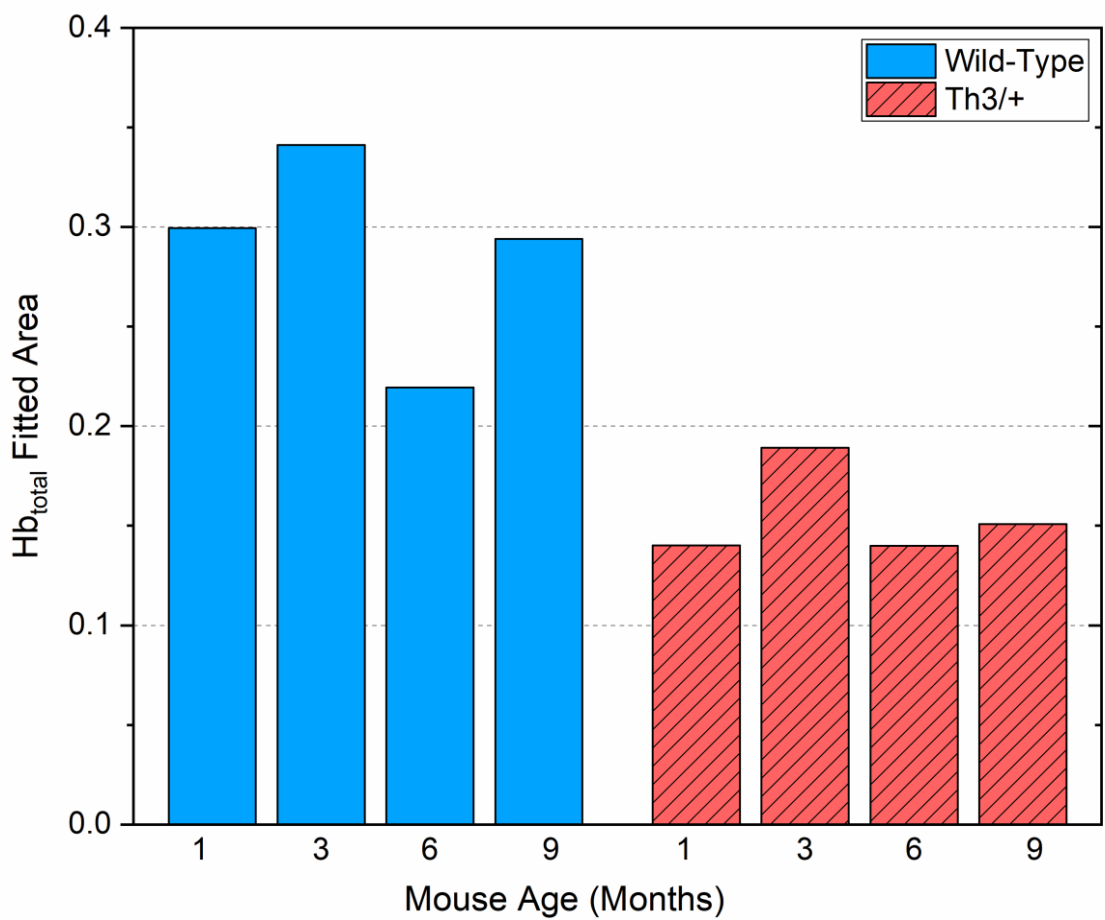

**Fig. 4** The fitted area of haemoglobin in the spectra from the blood samples of wild-type and th3/+ mice indicated, as a function of mice age. Hb is decreased in thalassaemic samples indicating anaemia

Ferritin-like iron was observed in the spectra from the blood samples of all th3/+ mice, in contrast to the wild-type ones where no such component was found. Unfortunately, it is not possible to determine from our methodology whether the ferritin-like iron was found inside the RBCs or in the serum since no cellular separation was performed for the reasons stated previously. Gardenghi et al (2007) showed that the th3/+ mice have serum iron levels of 20-30 mM which are slightly elevated compared to the serum iron levels in wild-type mice ( ~20 mM). In our measurements, ferritin-like iron in the spectra from blood samples of wild-type mice was untraceable. Thus, we can assume that the ferritin-like iron found in the blood of the th3/+ mice should exist mainly in the RBCs and not in the serum.

The percentage of the ferritin-like iron in the spectra from the blood of thalassaemic mice was found to decrease with age. The mouse at 1 month of age exhibited ferritin-like iron of ~ 14% while the mouse at 9 months of age ~ 6%. More interesting is that, the reduction of the ferritin-like iron with the age of the th3/+ mice seems to follow closely the corresponding reduction of the reticulocyte count, as presented by Gardenghi et al (2007). This is demonstrated in figure 5 where the percentages of the ferritin-like iron extracted from the MS spectra and the reticulocyte counts, measured by Gardenghi et al (2007), are plotted on the same figure.

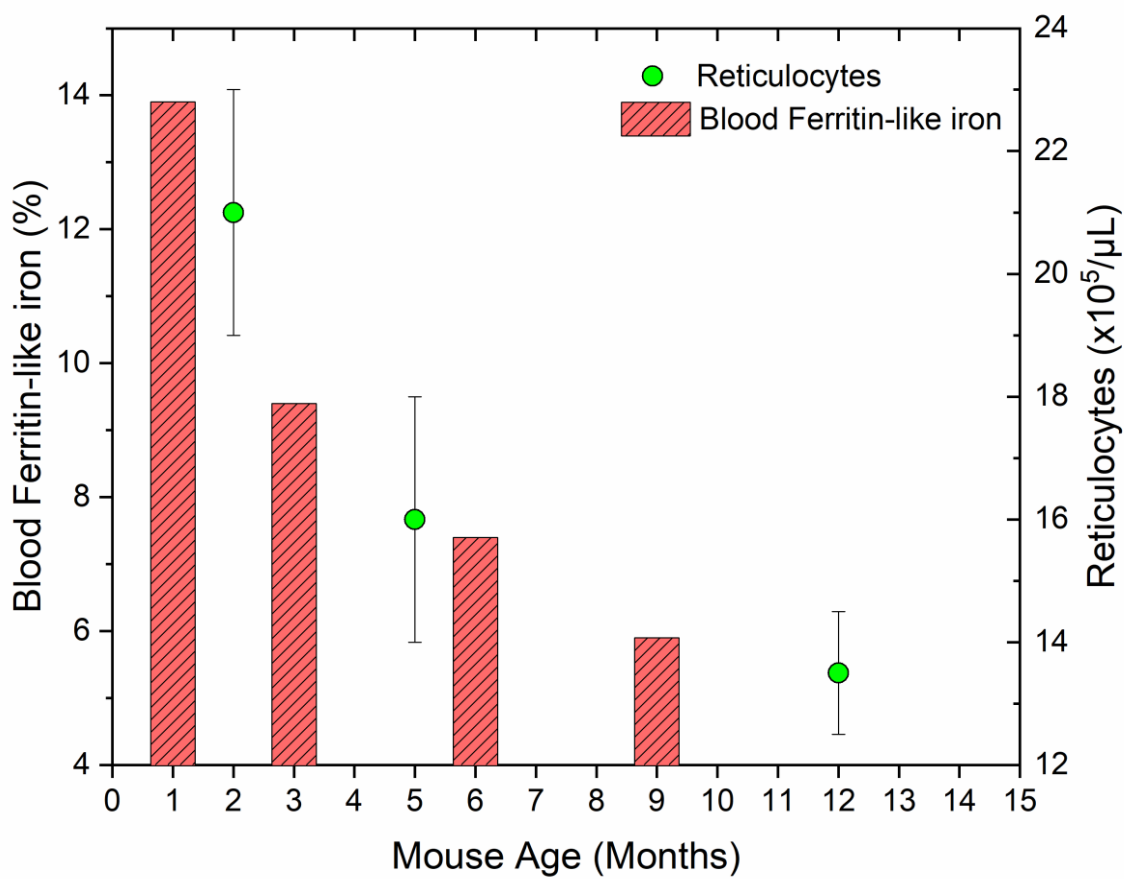

**Fig. 5** Percentage of ferritin-like iron in the Mössbauer spectra from the blood samples of th3/+ mice at 1, 3, 6 and 9 months of age. The reticulocyte count in th3/+ mice at 2, 5 and 12 months of age, as measured by Gardenghi et al. (2007) are also shown, for comparative purposes

These observations indicate that most of the ferritin-like iron found in the blood samples of the thalassaemic mice may be present in the reticulocytes. Thus, RBCs and more specifically the reticulocytes of thalassaemia patients might be possible carriers of a significant amount of iron. Note that the role of ferritin in the erythroid cells is still unclear. Ferritin has been suggested to act as an intermediate for haem synthesis in erythroid cells, but this has not yet been proven (Ponka et al. 2013). On the contrary, strong evidence has been provided that ferritin is not involved in hemoglobinization from a study by Darshan et al. (2009).

Ferritin-like iron was observed in the late 1970s, using MS, in the RBCs of thalassaemia intermedia and major patients (Bauminger et al. 1979), but this study was performed at a time that iron chelation therapy was not common and iron overload in the patients was extremely high. Increased amounts of liver-type ferritin was observed in the 80s (Piperno et al. 1984) by radioimmunoassay in the erythrocytes of patients with β-thalassaemia trait, thalassaemia intermedia and Cooley's disease (β-thalassaemia major). In a more recent study, Jiang et al. (1994) did not observe any ferritin-like iron in the RBCs of thalassaemia major patients that underwent iron chelation therapy with deferoxamine.

**Conclusions**

$^{57}$Fe enriched mice diet was prepared, and wild-type C57BL/6 and thalassaemic th3/+ mice were enriched through gastrointestinal absorption to characterize in more detail the iron complexes present

in the MS spectra. The enrichment method was validated by comparing, the MS spectra from the blood, liver and spleen of a non-enriched and a $^{57}$Fe enriched wild-type mouse at 1 month of age.

MS spectra from blood samples of wild-type and th3/+ $^{57}$Fe enriched mice at 1, 3, 6, and 9 months of age, were obtained at 80K. The spectra from the wild-type and th3/+ mice were compared in order to identify any differences between the healthy and thalassaemic mice, as well as any age related differences.

The spectra from the blood samples of the wild-type mice showed only the presence of oxyhaemoglobin, which was fitted by two Lorentzian sub doublets with a fixed 1:1 ratio, representing the α- and β-chains of haemoglobin. The average values from the fits of the wild-type samples were used subsequently to fit the oxyhaemoglobin's doublet of the th3/+ spectra, which showed an increased Hb-α/Hb-β ratio, probably due to excess α-globin chains.

The main difference between the blood sample spectra from the wild-type and th3/+ mice is that, for all mice ages studied, a significant amount of ferritin-like iron was observed in the thalassaemic samples but not in the wild-types ones. The observed ferritin-like iron in the blood of the th3/+ mice cannot be explained by an increase in the serum iron levels due to thalassaemia, as Gardenghi et al. (2007) showed that the serum iron levels in the th3/+ mice were elevated only slightly than the corresponding ones in wild-type mice. Thus, the ferritin-like iron identified in our study is likely to be located mainly in the RBCs of the mice.

Further, the percentage of the ferritin-like iron in the MS spectra from thalassaemic mice was found to decrease with the age of the mice. A similar decrease was observed in the reticulocyte count of the thalassaemic mice by Gardenghi et al. (2007). The source of this ferritin-like iron should be further investigated as the RBCs might be a hidden store of a significant amount of iron for thalassaemic patients. This might be of uppermost importance for the patients since erythrocyte ferritin is not utilized as a standard marker for the evaluation and management of thalassaemia treatment.

In a follow-up study, MS spectra from samples of various organs of $^{57}$Fe-enriched wild-type and th3/+ mice will be analyzed and compared to study the changes in iron accumulation in the organs of thalassaemic mice as a function of their age. This will provide a deeper insight into the iron complexes stored in the body due to thalassaemia.

**Acknowledgments:** We gratefully acknowledge the personnel of the Transgenic Mouse Facility of the Cyprus Institute of Neurology and Genetics for the planning of the mouse breeding and mouse feed preparation guidance.

**Ethical Approval:** All applicable international, national and/or institutional guidelines for the care and use of animals were followed.